\documentclass[showkeys,nofootinbib,prd]{revtex4}

\usepackage{amssymb}
\usepackage{amsmath}
 \usepackage{amsthm}
 \usepackage{graphicx}

\newtheorem{definition}{Definition}
\usepackage{hyperref}

\begin{document}

\title{Black to white hole transition as a change of the topology of the event horizon}
\author{Mattia Villani}
\affiliation{University of Urbino Carlo Bo, Department of Pure and Applied Sciences (DiSPeA), Via Santa Chiara, 27, Urbino (PU), 61029, Italy}
\email{mattia.villani@uniurb.it}

\begin{abstract}
We prove that the black to white hole transition theorized in several papers can be described as a change in the topology of the event horizon. We also show, using the theory of cobordism due to Milnor and Wallace, how to obtain the full manifold containing the transition.
\end{abstract}


\maketitle



\section{Introduction}
\label{sec:intro}
The possibility that black holes (BH) might explode as white holes (WH) was first suggested in \cite{planck1}. This was an application to Loop Quantum Gravity of a Loop Quantum Cosmology result \cite{lqc} (see also \cite{bojo}), according to which it is possible to resolve the initial Big Bang singularity, but the resulting Universe is cyclic. In the case of black holes, the idea is that when the density of the collapsing matter reaches values comparable to the Planck density, yet unknown quantum effects reverse the collapse, thus leading to an explosion. For a distant observer, this would be a black-to-white hole transition. In the reference frame of the collapsing matter this happens quickly, but a distant observer, because of the large time dilation, would see a stable black hole for millions of years. This black-to-white hole transition was explored in several papers \cite{planck2,planck3,planck4,planck5}. \cite{planck5}, in particular, shows that the probability of the explosion as a function of time is given by a decaying exponential with a time constant proportional to the square of the mass of the BH. This time constant is smaller than the evaporation one, which is proportional to the cube of the mass of the BH, and it suggests the possibility that lunar mass black holes have the largest probability to explode in this period of the life of the Universe. In other papers, such as \cite{barrau1,barrau2}, the authors have searched for possible observational signatures.

The main idea suggested in \cite{planck2,planck3} for describing WH is to join a copy of the Schwarzschild spacetime and its time-reversed along the singularity. Here we suggest another approach: we show that the explosion of a BH entails a change in the topology of its event horizon and try to describe the transition using the method of cobordism. We do not attempt to describe the (Quantum Gravity) process that triggers the transition; we can say that our treatment is classical.

This paper is organized as follows: in Section \ref{sec:why} we prove that a topology change indeed occurs at the level of the event horizon; in Section \ref{sec:topo} we calculate the Euler characteristic of the event horizon of a WH; in Section \ref{sec:mani} we use cobordism to describe the topology change; finally, in Section \ref{sec:concl}, we conclude our exposition.

\section{Why a topology change}
\label{sec:why}
A WH is a BH with the time coordinate reversed. We show that a black-to-white hole transition can occur if the topology of the event horizon changes.

We follow \cite{hawk,fr,gour,heus} and calculate the Euler characteristic according to the formula:
\begin{equation}\label{eq:chi}
    \chi=\int dS\, \left( R-2R_{\mu\nu\alpha\beta}l^\mu n^\nu l^\alpha n^\beta \right)+\int dS\left( 4\,\kappa_0\,T_{\mu\nu}l^\mu n^\nu \right)
\end{equation}
where $l^\mu$ is the generator of the event horizon and $n^\mu$ is a null vector such that $l^\mu n_\mu=1$ while $T_{\mu\nu}$ is the stress-energy tensor. The first integral can be rewritten as follows:
\begin{equation}
    R-2R_{\mu\nu\alpha\beta}l^\mu n^\nu l^\alpha n^\beta=-\dfrac{d\theta}{ds}-\beta^\mu_\sigma \beta^\nu_\rho\,\nabla_\mu n^\rho \nabla_\nu k^\sigma+p_\mu p^\mu-d^\dagger p
\end{equation}
where $\theta$ is the expansion of the null geodetic congruence and 
\begin{align}
    &\beta_{\mu\nu}=g_{\mu\nu}+l_\mu n_\nu+l_\nu n_\mu,\\
    &p^\mu=-n^\sigma \beta^{\mu\nu}\,\nabla_\nu l_\sigma,\\
    &d^\dagger p=-\beta^{\mu\nu}\nabla_\mu p_\nu,
\end{align}
where $g_{\mu\nu}$ is the spacetime metric. Integrating the first term, one obtains:
\begin{equation}
    \int dS\, \left( R-2R_{\mu\nu\alpha\beta}l^\mu n^\nu l^\alpha n^\beta \right)=\int dS\left( -\dfrac{d\theta}{ds}+p_\mu p^\mu \right)
\end{equation}
In the above integral, the second term is always positive, since $p^\mu$ is space-like. In a BH, the first term is also positive, because of the null version of the Raychauduri equation. For a WH, however, the first term is negative, since the light rays diverge at the event horizon. Moreover, because of the energy conditions, the second integral in \eqref{eq:chi} is also positive. For a BH, as shown in \cite{hawk} $\chi=2$ (a sphere), however, for a WH because of the change in sign of the expansion term, the Euler characteristic will be in general different from 2, thus there is a change in the topology of the event horizon in a black-to-white hole transition. We next determine the Euler characteristic of the event horizon of the WH.

\section{Determining the Euler characteristic}
\label{sec:topo}
We follow \cite{hawk} and determine the Euler characteristic of the WH. The idea in \cite{hawk} is to deform the event horizon into the exterior region in such a way that the future outgoing null geodesic would converge; however, this is a contradiction for a BH, see Sections 2 and 4 of \cite{hawk} for a complete discussion. 

We introduce the future directed null vector $l^\mu$ tangent to the horizon and another null vector $n^\mu$ orthogonal to the horizon, such that $l^\mu n_\mu=1$, as above. We deform the horizon by moving it a geodesic distance $-w$ along $n^\mu$. This is described by the Newman-Penrose equation:
\begin{equation}
    \dfrac{d\rho}{dw}=\overline{\delta}\tau+(\overline{\beta}-\alpha)\,\tau-\overline{\tau}\tau-\psi_2-2\Lambda
\end{equation}
with standard notation \cite{pen}. Following \cite{hawk}, we arrive at his equation (5), which is:
\begin{equation}\label{eq:questa}
    -\int(\psi_2+2\Lambda)\,dS=\pi \chi-\int(\Phi_{11}+3\Lambda)\,dS
\end{equation}
with $\Phi_{11}+3\Lambda\geq0$ for standard matter. At this point, \cite{hawk} concludes that one must have $\chi=2$, otherwise one finds a contradiction. However, following his reasoning, in our case, we can prove that the RHS of equation \eqref{eq:questa} is negative and that $\chi=0$ without contradiction because we now have the time-reversed of the event horizon of a black hole and the singularity lies in its past, not in its future, and the convergence parameter $\rho$ will be positive, see Sections 2 and 4 of \cite{hawk}.

Finally, we notice that this, however, is only half of the story, since then the WH ejects all its mass and reduces to nothing. This is a second topological transition and the full evolution of the event horizon topology will be described by the following steps:
\begin{equation}
    S^2\rightarrow T^2\rightarrow \varnothing.
\end{equation}
The second transition, however, is easy to describe: it will be a sequence of time-reversed Kerr spacetimes with a smaller an smaller mass and event horizon, until the mass reduces tozero and the event horizon disappears; at the end, we are left with a Minkowski spacetime. We now discuss the first transition.

\section{The first transition}
\label{sec:mani}
We try to describe the topological transition using cobordism. 

\begin{figure}
    \centering
    \includegraphics[width=0.55\linewidth]{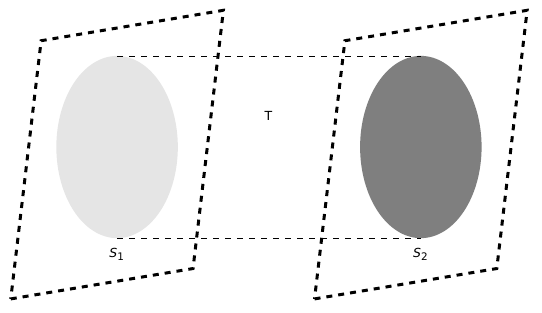}
    \caption{The idea behind non-compact manifold cobordism. The change in topology is confined in the cylinder $\mathcal{S}_1 \cup \mathcal{S}_2 \cup \mathcal{T}$.}\label{fig:idea}
\end{figure}

Usually, see \cite{wall}, cobordism is concerned with compact manifolds; here, however, we have non-compact ones. One can adapt the theory for compact manifolds to this case following \cite{topo,ger1,ger2}. If the topological transition is confined to a specific region of the spacetime $\hat{\mathcal{S}}_1$ and $\hat{\mathcal{S}}_2$, respectively in the initial and in the final manifolds, one introduces two spacelike spheres $\mathcal{S}_1$ and $\mathcal{S}_2$ surrounding the two regions, and in the interpolating manifold one introduces a timelike volume $\mathcal{T}$ joining them, see figure \ref{fig:idea}. The basic idea is to confine the topological transition to a compact region of the manifold.

We consider two surfaces which are slices of the Kerr spacetime and its time-reversed, respectively. The regions $\mathcal{S}_1$ and $\mathcal{S}_2$ are two spheres, $S^2$, containing the two event horizons. As shown above, the topology of the two event horizons is such that $\chi=2$ at the beginning and $\chi=0$ at the end. We need to determine the interpolating manifold $\hat{\mathcal{M}}$.

\begin{figure}
    \centering
    \includegraphics[width=0.55\linewidth]{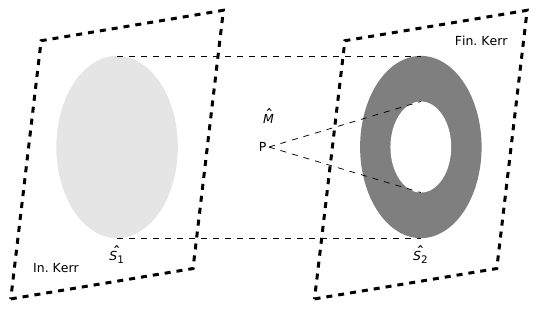}
    \caption{The initial and final manifolds with the interpolating manifold $\mathcal{M}$. $\mathcal{S}_1$ lies on the initial Kerr spacetime (In. Kerr), while $\mathcal{S}_2$ lies in the final Kerr spacetime (Fin. Kerr). $P$ is the critical point.}\label{fig:mani}
\end{figure}

We can define a Morse vector field $v$ \emph{entering} in $\hat{\mathcal{S}}_1$ and \emph{exiting} from  $\hat{\mathcal{S}}_2$. Following \cite{sor}, we have that the index of $v$ $\text{ind}(v)$ is such that:
\begin{equation}
    2\,\text{ind}(v)=\chi(\hat{\mathcal{S}}_2)-\chi(\hat{\mathcal{S}}_1)=-2
\end{equation}
Thus, the corresponding Morse function will have one non-singular critical point $p$ with index $-1$ and we have the identity:
\begin{equation}
    f(x)=f(p)-(x_1)^2+(x_2)^2+(x_3)^2+(x_4)^2.
\end{equation}
The way to construct $\hat{\mathcal{M}}$ was given in \cite{wall,yod1,yod2}: one has to carry out a series of \emph{spherical modifications} of $\hat{\mathcal{S}}_1$ that transform it into $\hat{\mathcal{S}}_2$. Spherical modifications are described by the following:
\begin{definition}[Spherical modification]
    Given a $n-1$ dimensional manifold, we remove from it  the set $D^{n-\lambda}\times S^{\lambda-1}$. This leaves a boundary $S^{n-1-\lambda}\times S^{\lambda-1}$. This boundary is identified with the boundary of the set $S^{n-1-\lambda}\times D^\lambda$. Such modifications are indicated with $(\lambda-1,n-1-\lambda)$.
\end{definition}
In our case, only one of such modifications is needed. This modification is of type $(0,2)$ with $\lambda=1$. The manifold $\mathcal{M}$ is reported in figure \ref{fig:mani}.

As discussed in \cite{yod2}, the critical point is problematic, since here the Morse vector field is zero, and no Lorentzian structure can be imposed on the manifold since the Lorentzian metric would be singular. However, this point can be eliminated with the surgery described in \cite{yod1,mil}: the idea is to cut a volume $D$ around the critical point leaving it with a $S^3$ boundary; then we do the same on another manifold $\mathcal{N}$ which has the same dimension as $\mathcal{M}$; we then paste $\mathcal{M}-D$ and $\mathcal{N}-D$ at the $S^3$ boundary thus obtaining the manifold $\hat{\mathcal{M}}^\prime$: if the manifold $\mathcal{N}$ has been chosen appropriately, one can define a non-singular vector field $v$ and $\hat{\mathcal{M}}^\prime$ will be differentiable. In this way, we have obtained a Lorentzian manifold $\hat{\mathcal{M}}^\prime$ which interpolates between our two initial manifolds.

\section{Discussion and conclusion}
\label{sec:concl}
We have proved that the black-to-white hole transition is a change in the topology of the event horizon, which passes from $S^2$ to $T^2$. The subsequent loss of mass from the WH is another topological transition of the type $T^2\rightarrow\varnothing$, so the final spacetime is Minkowski.

\begin{figure}
    \centering
    \includegraphics[width=0.95\linewidth]{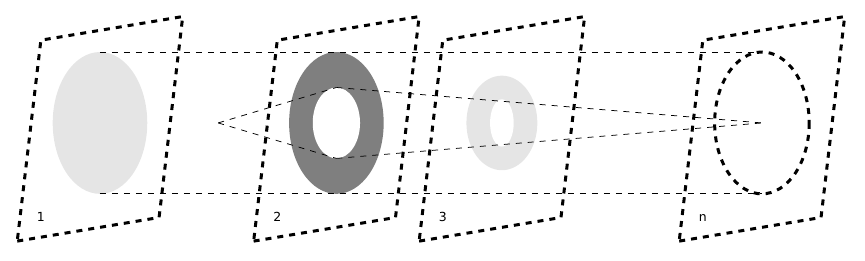}
    \caption{The complete $S^2\rightarrow T^2\rightarrow\varnothing$ transition. The manifold $1$ is a Kerr spacetime; manifolds $2,3,\dots$ are time-reversed Kerr spacetimes, while $n$ is a Minkowski spacetime.}\label{fig:tot}
\end{figure}

We have described the first topological transition using cobordism between a Kerr spacetime and its time-reversed, and we were able to obtain the interpolating manifold. As said above, the second topological transition can be described by a sequence of time-reversed Kerr spacetimes with smaller and smaller mass down to $M=0$ when the event horizon disappears and the spacetime becomes Minkowski. This full process is described in figure \ref{fig:tot}. We see that the manifold that describes the complete transition $S^2\rightarrow T^2\rightarrow\varnothing$ must have a hole with $S^3$ boundary, thus one can conclude that the first homotpy groups of this manifold are:
\begin{equation}
    \pi_1=\varnothing, \quad \pi_2=\varnothing, \quad \pi_3=\mathbb{Z}.
\end{equation}

We did not attempt to find the actual mechanism which leads to the first transition, which must be looked for in a Quantum Gravity processes. We leave this for a future publication.

\end{document}